\documentclass[twoside]{article}
\usepackage{qic,epsfig}
\usepackage{color,soul}
\usepackage{subfig}
\usepackage{float}
\usepackage{graphicx}
\usepackage[dvipsnames]{xcolor}
\usepackage{dcolumn}
\usepackage{bm}
\usepackage{braket}
\usepackage{amsmath}
\usepackage{siunitx}
\newcommand{\SERF}{0.10}
\newcommand{\SERS}{0.20}
\newcommand{\SERL}{0.37}
\newcommand{\TERF}{0.11}
\newcommand{\TERS}{0.33}
\newcommand{\TERL}{0.52}

\renewcommand{\thefootnote}{\fnsymbol{footnote}}  

\begin{document}
\setlength{\textheight}{8.0truein}    

\runninghead{Rapid and Robust generation of Einstein-–Podolsky–-Rosen pairs with Spin Chains}
            {Kieran N. Wilkinson, Marta P. Estarellas, Timothy P. Spiller, Irene D'Amico}

\normalsize\textlineskip
\thispagestyle{empty}
\setcounter{page}{1}


\vspace*{0.88truein}

\alphfootnote

\fpage{1}

\centerline{\bf
\uppercase{Rapid and Robust generation}}
\vspace*{0.035truein}
\centerline{\bf  \uppercase{of Einstein--Podolsky--Rosen pairs with Spin Chains}}
\vspace*{0.37truein}
\centerline{\footnotesize
KIERAN N. WILKINSON}
\vspace*{0.015truein}
\centerline{\footnotesize\it Department of Physics, University
of York, YO10 5DD}
\baselineskip=10pt
\centerline{\footnotesize\it York, United Kingdom}

\vspace*{10pt}
\centerline{\footnotesize 
MARTA P. ESTARELLAS}
\vspace*{0.015truein}
\centerline{\footnotesize\it Department of Physics, University
of York, YO10 5DD}
\baselineskip=10pt
\centerline{\footnotesize\it York, United Kingdom}

\vspace*{10pt}
\centerline{\footnotesize 
TIMOTHY P. SPILLER}
\vspace*{0.015truein}
\centerline{\footnotesize\it Department of Physics, University
of York, YO10 5DD}
\baselineskip=10pt
\centerline{\footnotesize\it York, United Kingdom}

\vspace*{10pt}
\centerline{\footnotesize 
IRENE D'AMICO}
\vspace*{0.015truein}
\centerline{\footnotesize\it Department of Physics, University
of York, YO10 5DD}
\baselineskip=10pt
\centerline{\footnotesize\it York, United Kingdom}

\vspace*{0.225truein}
\centering\today
\vspace*{0.21truein}

\abstract{
We investigate the ability of dimerized spin chains with defects to generate EPR pairs to very high fidelity through their natural dynamics. We propose two protocols based on different initializations of the system, which yield the same maximally entangled Bell state after a characteristic time. This entangling time can be varied through engineering the weak/strong couplings' ratio of the chain, with larger values giving rise to an exponentially faster quantum entangling operation. We demonstrate that there is a set of characteristic values of the coupling, for which the entanglement generated remains extremely high. We investigate the robustness of both protocols to diagonal and off-diagonal disorder. Our results demonstrate extremely strong robustness to both perturbation types, up to strength of 50\% of the weak coupling. Robustness to disorder can be further enhanced by increasing the coupling ratio. The combination of these properties makes the use of our proposed device suitable for the rapid and robust generation of Bell states in quantum information processing applications.}

\vspace*{10pt}

\vspace*{3pt}

\vspace*{1pt}\textlineskip    
\newpage
\section{Introduction}        
The efficient and reliable production of entangled states is an area of particular importance in quantum technologies due to the relevance of such states as a resource in numerous applications (e.g. one-way quantum computation architectures \cite{raussendorf2001,walther2005}, secure quantum key distribution schemes \cite{jennewein2000quantum, bennett1992quantum} and teleportation protocols \cite{bouwmeester1997,braunstein1998}). EPR pairs are pairs of qubits in the form of a Bell state and exhibit the simplest example of maximal bipartite entanglement \cite{einstein1935}. Their importance as an entanglement resource is well established, hence various protocols for generating such pairs have been proposed \cite{kwiat1995,hucul2015}. While most of these proposals involve the use of photons and optical implementations, some quantum architecture schemes may require a device able to generate such entangled states on a microscopic scale or within solid-state architectures. Therefore devices able to create and distribute such states on demand, by also linking quantum registers within solid-state quantum computer architectures are desirable.  In this paper we propose two different protocols to accomplish this using spin chains.

Spin chains have been widely studied as a means to carry and distribute quantum information over short distances and their significance as a reliable quantum bus has been established over the past decades \cite{bose2007,niko2004,kay2010}. Other applications such as their use as entanglers \cite{yung2005,estarellas2017robust,srivinasa2007,Venuti2006,Venuti2007,Giampaolo2009,Banchi2011,niko2004_b,nikoBook} or to localise states in protected ways \cite{estarellas2016} have also been proposed. Among the advantages of spin chains as quantum devices is that they can be implemented through many physical systems. Examples include electrons and excitons trapped in nanostructures \cite{niko2004,damico2007}, nanometer scale magnetic particles \cite{tejada2001}, strings of fullerenes \cite{twamley2003} and, more generally, any physical system formed by a string of coupled two-level quantum subsystems.

In this paper we consider spin chains than can be represented using the XY Heisenberg model. With the use of such system we propose two protocols to generate EPR states that use two different system initializations. In the first, the injection of a single-excitation at the center allows us to generate and distribute the entanglement of a Bell state between two distant parties. The second one, spanning the two-excitation subspace, entangles two qubits that were initially distant and distributes the Bell state back to these two initial parties. We then present detailed results of the dynamics of the system both from the numerical and analytical perspectives for both protocols. We also account for imperfections in the fabrication of the device by exploring the effects of diagonal and non-diagonal static disorder on the overall protocols. For both the ideal and perturbed cases, we study in detail the dependence of the entangling protocol with the spin chain characteristics, showing the optimal scenarios. We finally compare the entangling times that could be achieved by different possible physical implementations of the protocol. Our results demonstrate that our proposed system offers a robust and rapid solid state alternative to generate Bell states, applicable to a wide range of physical systems.

\section{The Model}
\noindent
The system considered here is a $N$=7 sites linear spin chain tuned to have an `ABC' configuration, which consists of staggered weak ($\delta$) and strong ($\Delta$) couplings, in a distribution such that there are three sites (named A, B and C) weakly coupled to the rest of the chain, as shown in Fig.\ \ref{fig:protocol}. The relevance of these three sites will be explained later.

The spin chain can be described by the following time-independent Hamiltonian,
\begin{eqnarray}
\label{hami}
{\cal{H}} = \sum_{i=1}^{N}\epsilon_{i}|1\rangle \langle 1|_{i} + \sum_{i=1}^{N-1} J_{i,i+1}\big[ |1\rangle \langle 0|_{i} \otimes |0\rangle \langle 1|_{i+1} + h.c.\big],
\end{eqnarray}
with $J_{i,i+1}$ equal to either $\Delta$ or $\delta$ depending on the site. We will refer frequently to the coupling ratio, $\delta/\Delta$, as its value affects the properties we are interested in. Without loss of generality, we set $\Delta=1$ in all cases so that the weak coupling strength will determine the coupling ratio. The first term in the Hamiltonian describes the on-site energies, $\epsilon_{i}$, which will be considered to be site independent (set to zero for convenience) until we include the effect of random diagonal disorder at a later stage. In our encoding we define a single excitation $|1\rangle$ as an ``up" spin in a system initially prepared to have all the spins ``down", $|0\rangle$. In previous literature \cite{estarellas2017robust,estarellas2016,huo2008,Almeida2016,Venuti2007} it has been demonstrated that related dimerized chains have high fidelity quantum state transfer (QST) properties, which we will exploit in this investigation. 

\subsection{Entanglement generation protocols}

\begin{figure}[h!]
    \centering
    \includegraphics[width=0.55\textwidth]{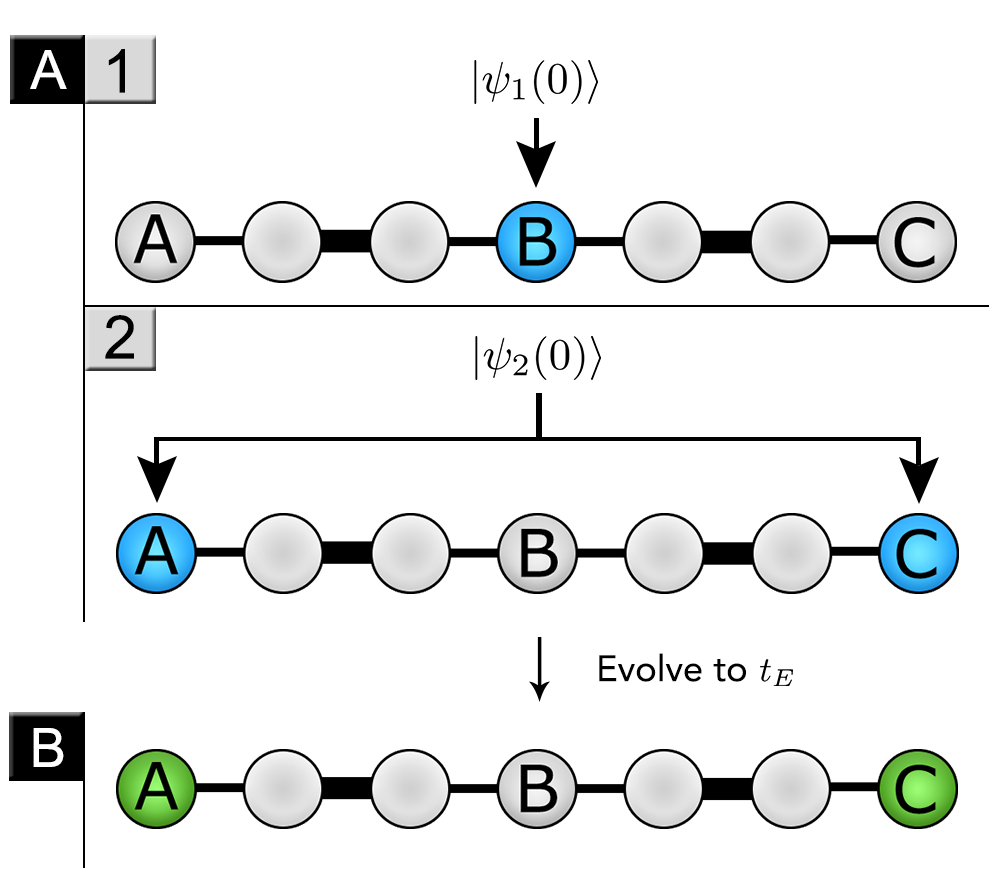}
    \caption{Entangling protocols: (A) An ABC spin chain consisting of $N=7$ sites (with thick lines representing $\Delta$ and thin lines representing $\delta$) is prepared by injecting either (i): the single-excitation state, $\ket{\psi_1(0)}$ (Eq.\ (\ref{is1})) or (ii): the two-excitation state, $\ket{\psi_2(0)}$ (Eq.\ (\ref{is2})). (B) Qubits A and C are entangled (green) after time evolution of the natural system dynamics; $t_E$ is the time at which maximum entanglement is achieved.}
    \label{fig:protocol}
\end{figure}

We now introduce our two protocols for generating entangled states with ABC spin chains. Both protocols are comprised of two steps: firstly we inject a single product state into the system at $t=0$. We consider the initial injection of one of the following two states:

\begin{romanlist}
 \item The single-excitation state injected at site B, $\ket{\psi_1(0)}$, given by
\begin{equation}
    \ket{\psi_1(0)} = |1\rangle_B\otimes|0\rangle_{\text{rest-of-chain}}.
    \label{is1}
\end{equation}
 \item The two-excitation state with a simultaneous injection at site A and C, $\ket{\psi_2(0)}$, given by
\begin{equation}
    \ket{\psi_2(0)}= |1\rangle_A\otimes|0\rangle_{\text{rest-of-chain}}\otimes|1\rangle_C.
    \label{is2}
\end{equation}

\end{romanlist}
We then allow the chosen initial state to evolve through the natural dynamics of the system, and observe the formation of an entangled state between qubits A and C at a later characteristic time $t_E$. The protocols are outlined graphically in Fig.\ \ref{fig:protocol}. For convenience, we will henceforth refer to the methods requiring the initial injection of the single- and two-excitation states as the single- and two-excitation protocols, respectively. 

In order to quantify the bipartite entanglement between sites A and C, we use the entanglement of formation, EOF, a bipartite measure of entanglement for both pure and mixed states \cite{wootters1998}. For the pair of qubits A and C, the EOF is defined by,
\begin{equation}
EOF_{AC}=\xi(C_{AC}),
\end{equation}
where $\xi(C_{AC})=h(\frac{1+\sqrt{1-\tau}}{2})$ and $h=-x\log_2x-(1-x)\log_2(1-x)$ \cite{wootters2001}.
We can compute this value by obtaining the square roots of the four eigenvalues, $\lambda_i=\sqrt{\varepsilon_i}$, of the matrix $\rho_{AC}\widetilde{\rho_{AC}}$ (with $\rho_{AC}$ being the reduced density matrix of sites A and C, and $\widetilde{\rho_{AC}}=(\sigma^A_y\otimes\sigma^C_y)\rho^*_{AC}(\sigma^A_y\otimes\sigma^C_y)$), and arranging these $\lambda_{i}$ in decreasing order. Then $\tau$ is obtained as

\begin{equation}
\tau=(\max(\lambda_1-\lambda_2-\lambda_3-\lambda_4,0))^2.
\end{equation}

In the $\delta/\Delta\ll1$ limit, the full 7-site ABC spin chain system can be approximated by a smaller `toy model', in which we consider only the three sites A, B and C, equally coupled, such that $J_{AB}=J_{BC}=\eta$, where $\eta$ is the effective coupling induced by the presence of the connecting dimers. The Hamiltonian of this system is then given by
\begin{equation}
H=
\begin{pmatrix}
0 & \eta & 0 \\
\eta & 0 & \eta \\
0 & \eta & 0
\end{pmatrix} \;,
\label{hamiT}
\end{equation}
and the simplicity of its form allows us to easily obtain its eigenstates. The time-evolution of these eigenstates will be used as a base to obtain an analytical approximation of the dynamics of the single- and two-excitation protocols. 

\begin{figure}[ht!]
    \centering
    \resizebox{0.8\textwidth}{!}{
\subfloat{{\includegraphics{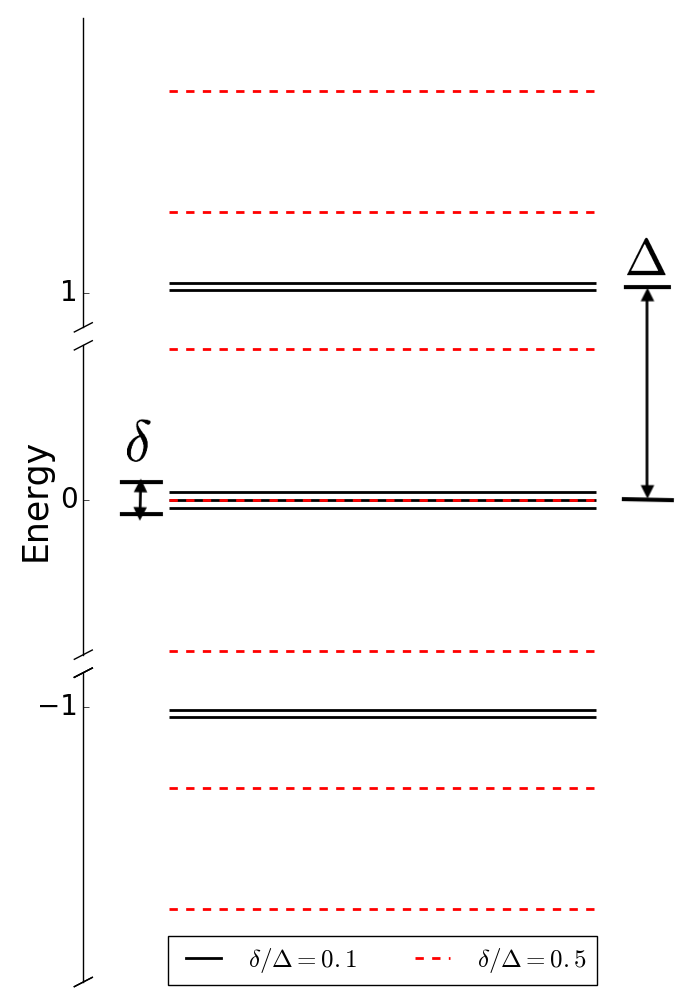}}}%
    \qquad
\subfloat{{\includegraphics{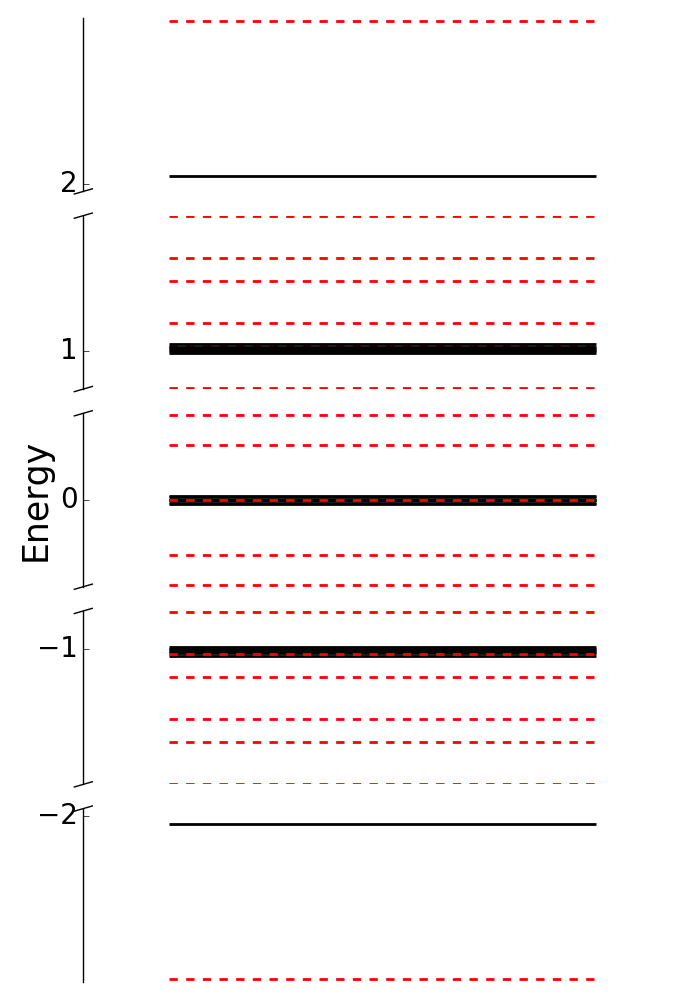} }}%
    }
    \caption{Comparison of the energy spectra of our 7-site ABC spin chain with one (l.h.s) and two (r.h.s) excitations. The black and red-dashed lines represent the energy levels for a system with a coupling ratio $\delta/\Delta=0.1$ and $\delta/\Delta=0.5$, respectively. It can be observed from this figure that both the increasing of the coupling ratio and the excitation number rises the complexity of the spectrum. Because of this, the trimer model is a worse approximation of the three relevant states that sit in the energy gap.}%
\label{spectra}       
\end{figure}

It is important to note that when the coupling ratio and the number of excitations increases the accuracy of this approximation to describe the dynamics falls off. This can be understood by inspecting the energy spectrum as shown in Fig.\ \ref{spectra}. The trimer approximation is valid for low coupling ratios because the dimerization of the chain opens a large energy gap $\propto\Delta$. At the center of this gap, there are three states. For values of $\Delta\gg\delta$, the dynamics involving such three states can be considered independently of the rest of the spectrum. However, as the $\delta/\Delta$ increases, the gap shrinks as the separation between the state of each band and central states depend on $\delta$. Therefore the states from the bulk are now closer in energy to the trimer-like states. This will unavoidably have an influence on the dynamics causing bigger secondary oscillations as the energy difference gets smaller. This influence is even more abrupt when considering the two excitation sector due to the interplay of other states and a more complicated energy spectrum (see r.h.s panel of Fig.\ \ref{spectra}). 

In most of the results we will present, we simulate the system dynamics over a time window the size of a single ``fidelity period'', $t_F$, which is defined as the time taken for the chain to regain (or almost regain) its initial state after $t=0$. We calculate $t_F$ using our `toy model'. In general the fidelity, $F$, is a measure of the overlap of a desired or target state of the chain $\psi_{tar}$ with the state of the chain at a given time $\psi(t)$, and is given by
\begin{equation}
	F = |\braket{\psi_{tar}|\psi(t)}|^2.
\end{equation}
A value of $F=1$ represents complete overlap of the two states. As will be seen, if  for our system the dynamics and the fidelity against the initial state, $\psi_{tar}=\psi(0)$, are calculated for an oscillation period $t_F$, the desired state---a maximally entangled state---will occur during this oscillation, at time $t_{E}$ when the fidelity against the initial state approaches zero. For our trimer model, it can be shown analytically that the fidelity period is given by $t_F=\pi/(\sqrt{2}\eta)$ (see Appendix A), and we can relate this formalism back to our 7 site ABC chain by finding the value of the effective coupling strength between the key sites, which is given by 

\begin{equation}
\eta=\frac{\Delta}{2}\sqrt{1+3\bigg(\frac{\delta}{\Delta}\bigg)^2-\sqrt{1+6\bigg(\frac{\delta}{\Delta}\bigg)^2+\bigg(\frac{\delta}{\Delta}\bigg)^4}}. 
\label{eta}
\end{equation}

\noindent This follows directly from the separations in energy between the three states sitting in the energy gap, which are the main ones playing a role in the dynamics. These separations are influenced by the coupling ratio and, therefore, the sites that sit between A-B and B-C. The effective trimer coupling $\eta$ is obtained by calculating the energy spectrum of the full Hamiltonian written in terms of $\Delta$ and $\delta$ and relating it to the trimer spectrum, as it is shown in Appendix \ref{appb}.  

We expect $t_E$ to be close to $t_F/2$ - the middle of the window we consider: this is when, for the `toy model', the fidelity of the initial state falls to zero, accompanied by an EoF peak close to unity due to the emergence of the now entangled state. Despite the decrease in accuracy of our approximation for the dynamics for large coupling ratios, the trimer approximation still holds well to calculate these time scales.

The chain length in this model can be increased by adding sets of four sites (two dimers, one either side of site $B$ to preserve the symmetry) and the system will still support the protocols presented here. Yet this chain growth would increase the time taken for entanglement creation, exponentially with chain length, due to the exponential decrease of $\eta$ with length. The important feature of the application proposed in this work is the robust creation of entanglement, therefore the scalability with chain length is less of an issue for this that it would be for applications with the chain acting as a quantum communication bus only.

\setcounter{footnote}{0}
\renewcommand{\thefootnote}{\alph{footnote}}

\section{Results}
\subsection{Single-excitation protocol}

\begin{figure}[h!]
    \centering
    \includegraphics[width=0.7\textwidth]{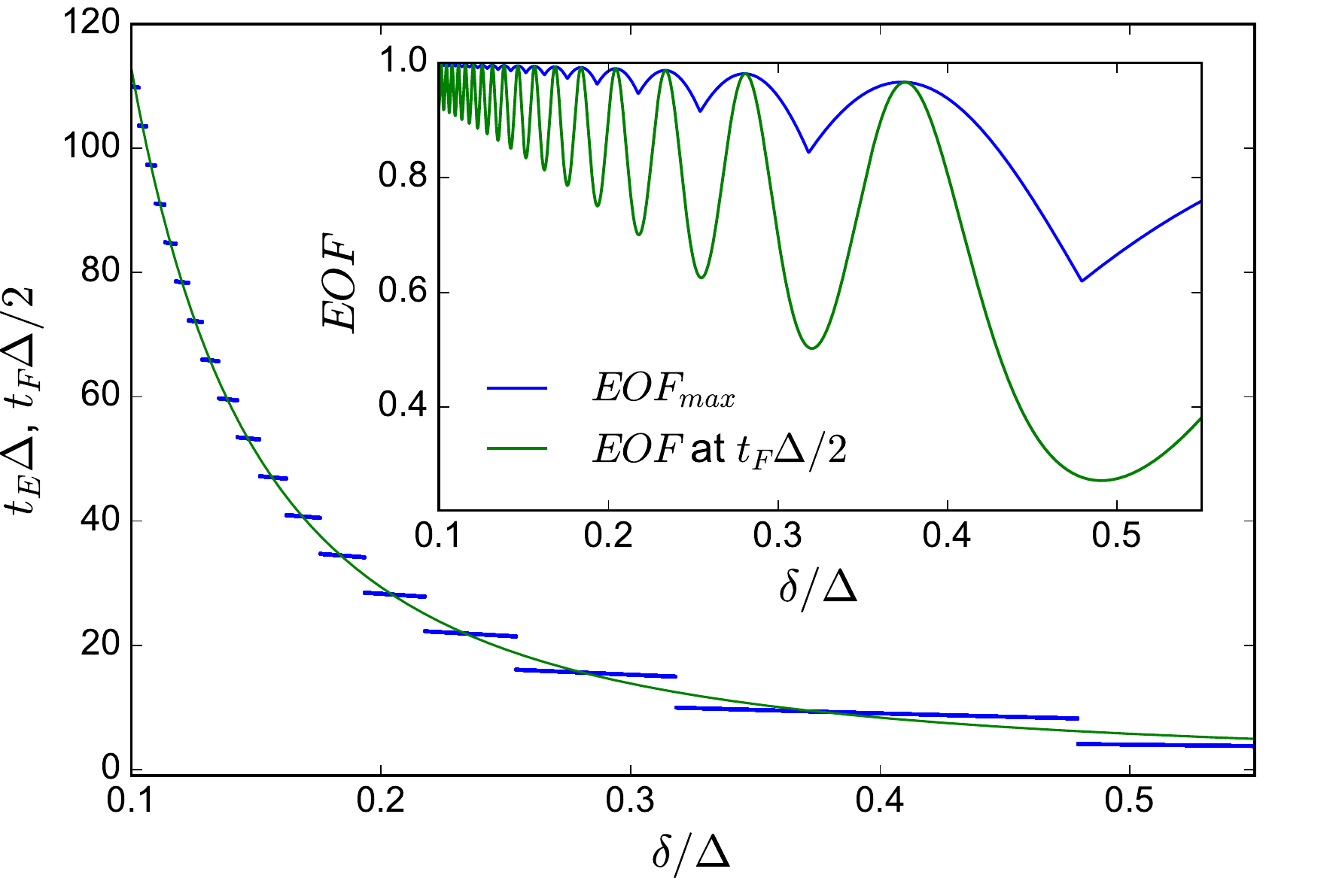}
    \caption{Variation of $t_E$ with coupling ratio (blue data points) for the single-excitation protocol with overlaid analytical formula for $t_F/2$ (green line). Inset: maximum EOF at $t_E$ (blue line) and EOF at the analytical $t_F/2$ (green line).}
    \label{maxEoFoneex}
\end{figure}

We first consider the coupling ratio dependence of both the time taken to achieve maximum entanglement ($t_E$), and the associated EOF value, which we denote as $EOF_{max}$. As shown by Fig.\ \ref{maxEoFoneex}, the overall trend is that the $t_E$ decreases exponentially with increasing coupling ratio. We observe, however, ranges of coupling ratios with approximately constant $t_E$ (shown as `flat' blue segments). Each of these ranges presents however only a single coupling ratio value for which the associated $t_E$ is approximately $t_F/2$ (green line). This feature is demonstrated in the inset where we observe that the $EOF_{max}$ profile is a sequence of ``arches'', each of them corresponding to one of the almost constant $t_E$ regions of the main panel. The coupling ratios for which $t_E=t_{F/2}$  coincide with the peak of each ``arch'' of the $EOF_{max}$ profile. Even considering the arches, $EOF_{max}$ between A and C remains very high for a large range of coupling ratios.

\begin{figure}[ht!]
\centering
\resizebox{0.7\textwidth}{!}{
  \includegraphics{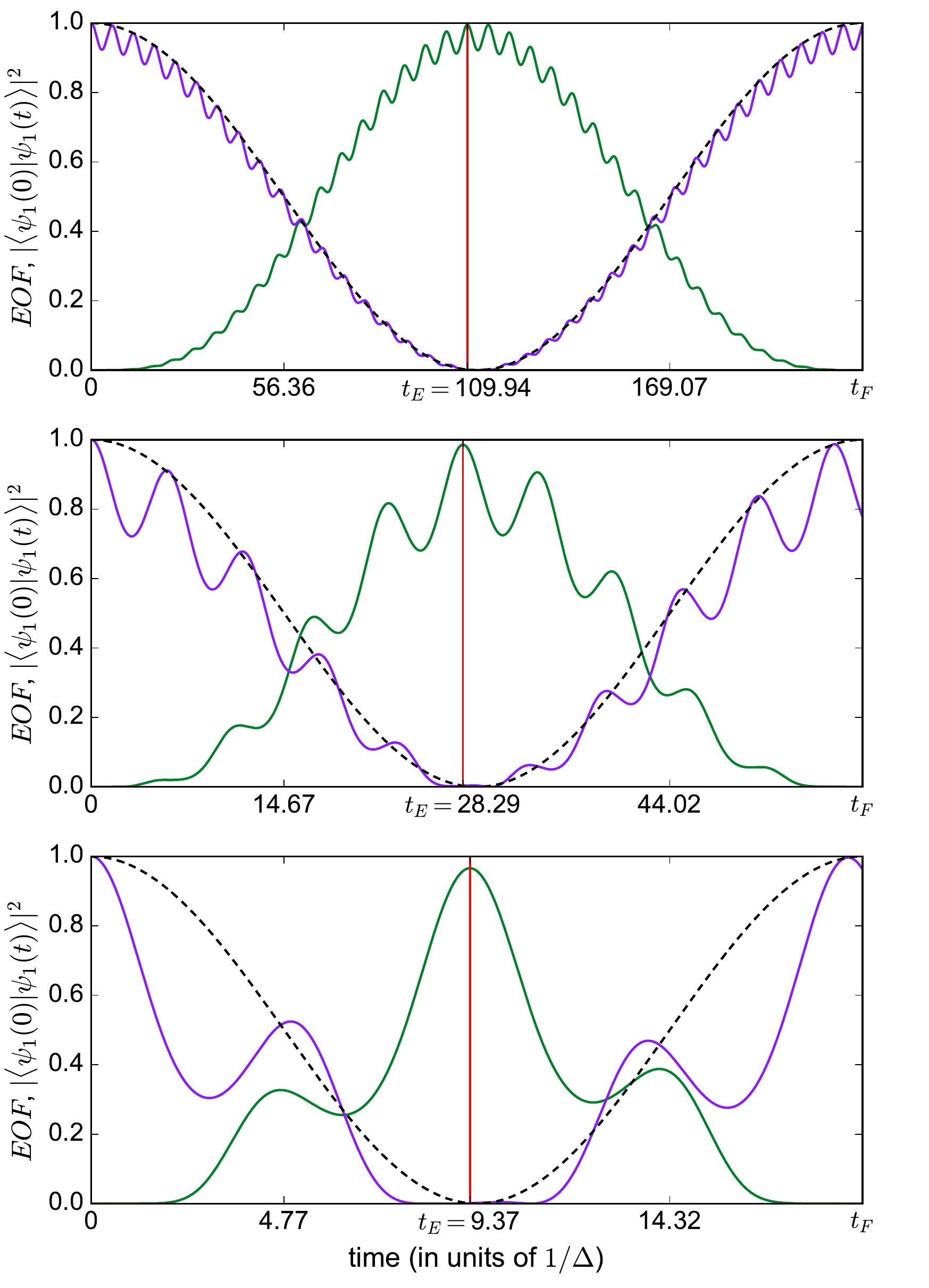}
}
\caption{Fidelity of the initial state (purple) and EOF (green) for the single-excitation protocol with ratios \SERF \ (top), \SERS \ (middle), and \SERL \ (bottom). The vertical red line denotes $t_E$, the time at which the maximum value of the EOF is obtained. The black dashed line is the analytically obtained fidelity for the trimer corresponding to each system.}
\label{panOneEx}       
\end{figure}

We now study in detail the three coupling ratios \SERF, \SERS\, and \SERL\ that correspond to local maxima of the $EOF_{max}$ profile (see Fig.\ \ref{maxEoFoneex}). For each ratio we simulate the system dynamics over a time window the size of a single analytic fidelity period $t_F$, wherein we consider both the fidelity of the initial state and the EOF between sites A and C. The results for each ratio are shown in Fig.\ \ref{panOneEx} with purple and green lines representing fidelity and EOF, respectively. We observe an EOF peak close to the center of the window (corresponding to $t_F/2$) with a fidelity of the initial state close to zero. We note that our analytical estimates for $t_F$ are accurate despite underlying secondary oscillations resulting from the presence of the additional sites other than just A, B and C in the chain. The amplitude of these oscillations increases significantly with coupling ratio as a result of the reduction in the energy difference of the couplings. At the same time, their frequency in units of $t_F^{-1}$ decreases, reducing to unity in the limit $\delta/\Delta\rightarrow1$. The arches of $EOF_{max}$ we observed in the inset of Fig.\ \ref{maxEoFoneex} are generated by the presence of this secondary frequency, each subsequent arch corresponding to one secondary period less within $t_F$. The variation of $EOF_{max}$ within each arch then tracks the maximum of a secondary period going in and out of phase with the maximum of the main period.

The state which we observe when the EOF reaches its maximum value is, up to small corrections (with these smaller the closer $EOF_{max}$ is to unity), a Bell state of the form
\begin{equation}
	\ket{\psi_1(t_E)}=\frac{1}{\sqrt{2}}\big(\ket{10}_{\text{AC}} + \ket{01}_{\text{AC}}\big)\otimes \ket{0}_{\text{rest-of-chain}}.
    \label{entang}
\end{equation}
We now use the aforementioned trimer model to better understand the underlying entanglement mechanism behind this protocol. We begin by considering the eigenstates of the reduced Hamiltonian Eq.\ (\ref{hamiT}), which can be obtained through exact diagonalization and are given by
	\begin{equation}
	|\phi_{-}\rangle=\frac{1}{2}
	\left(
	\begin{matrix}
	-1\\\sqrt{2}\\-1
	\end{matrix}
	\right),\quad
	|\phi_{0}\rangle=\frac{1}{\sqrt{2}}
	\left(
	\begin{matrix}
	1\\
	0\\
	-1\\
	\end{matrix}
	\right), \quad \text{and}\quad 
	|\phi_{+}\rangle=\frac{1}{2}
	\left(
	\begin{matrix}
	1\\
	\sqrt{2}\\
	1\\
	\end{matrix}
	\right),
	\label{eigenstates}
	\end{equation}
    with $|\phi_{-}\rangle$ having energy $E_{-}=-\sqrt{2} \eta$, $|\phi_{0}\rangle$ having energy $E_{0}=0$,  and $|\phi_{+}\rangle$ having energy $E_{+}=\sqrt{2} \eta$. In terms of these eigenstates, the state of the chain at $t=0$, can be decomposed as
   \begin{equation}
   |\Psi(t=0)\rangle=\frac{1}{\sqrt{2}}(|\phi_{+}\rangle+|\phi_{-}\rangle)
   \label{ISA}
   \end{equation}
and we can time-evolve each of the two constituent eigenstates, $\ket{\phi_{+}}$ and $\ket{\phi_{-}}$ using their respective propagators, $e^{-iE_+t}$ and $e^{-iE_-t}$. From the results presented in Fig.\ \ref{maxEoFoneex} and \ref{panOneEx}, EOF is maximum at $t_F/2$, hence using this time value, the state of the system can be expressed as
\begin{equation}
\bigg|\Psi\bigg(\frac{\pi}{2\sqrt{2}\eta}\bigg)\bigg\rangle=\frac{1}{\sqrt{2}}\big(e^{-iE_{+}\frac{\pi}{2\sqrt{2}\eta}}|\phi_{+}\rangle+e^{-iE_{-}\frac{\pi}{2\sqrt{2}\eta}}|\phi_{-}\rangle\big)=\frac{1}{\sqrt{2}}(-i|\phi_{+}\rangle+i|\phi_{-}\rangle).
\end{equation}
Using the explicit expression of the basis vectors, this yields a maximally entangled Bell state between sites A and C:

\begin{eqnarray}
\bigg|\Psi\bigg(\frac{\pi}{2\sqrt{2}\eta}\bigg)\bigg\rangle
=\frac{-i}{\sqrt{2}}
\left[
\begin{pmatrix}
0\\
0\\
1
\end{pmatrix}
+
\begin{pmatrix}
1\\
0\\
0
\end{pmatrix}
\right].
\label{bell}
\end{eqnarray}
We note that this entangled state is orthogonal to $\ket{\psi_1(0)}$, hence we expect the EOF to be maximum when the fidelity for the ABC chain is very small, and this is indeed observed in Fig.\ \ref{panOneEx}. To further explore this approximation, we can calculate analytically the fidelity of the trimer when the initial state is that of Eq.\ (\ref{ISA}). This result is presented with a black dashed line profile for the dynamics associated with the three ratios shown in Fig.\ \ref{panOneEx}. The approximation works well at low ratios but deviates considerably from the result for the full chain due to increasingly large secondary oscillations as the ratio increases. In all cases, however, we observe that the maxima and minima of the numeric and analytic results coincide closely; our analytical model predicts the key features of the system dynamics.

\subsection{Two-excitation protocol}
The single excitation protocol allows us to generate and distribute a Bell state with the convenience of having to initially interact only with a single site (B). However, some applications such as modular quantum processors proposals may need to generate this same entangled state with compliance and contribution of two distant parties or quantum registers \cite{Gualdi2011}. This is the motivation that leads us to propose the two-excitation protocol. We introduce it by demonstrating that an entangled state similar to that obtained through the single-excitation mechanism can be obtained through the injection of $\ket{\psi_2(0)}$. We do this by noting the particle-hole symmetry of the reduced trimer model and the fact that, for a three site system, one and two excitation pure states linked by this symmetry ($|100\rangle\rightarrow|011\rangle$, $|010\rangle\rightarrow|101\rangle$, and $|001\rangle\rightarrow|110\rangle$) are energetically indistinguishable as long as on site energies are set to zero. We can hence perform a swap operation between $\ket{0}$ and $\ket{1}$ states of the state calculated in Eq.\ (\ref{bell}) and obtain the particle-hole symmetric counterpart entangled state at $t_F/2$ for the two excitations injection, of the form:

\begin{equation}
\bigg|\Psi\bigg(\frac{\pi}{2\sqrt{2}\eta}\bigg)\bigg\rangle=
\frac{-i}{\sqrt{2}}
\left[
\begin{pmatrix}
1\\
1\\
0
\end{pmatrix}
+
\begin{pmatrix}
0\\
1\\
1
\end{pmatrix}
\right].
\end{equation}
We therefore expect injection and evolution of $\ket{\psi_2(0)}$ to yield the same maximally entangled Bell state between sites A and C, with the extra excitation sitting in the middle. This state indeed coincides with the one we numerically observe at $t_E$ for  the ABC chain, which was in fact the following:

\begin{eqnarray}
\ket{\psi_2(t_E)}=\frac{1}{\sqrt{2}}\big[\big(\ket{10}_\text{AC} + \ket{01}_\text{AC}\big)\otimes \ket{1}_\text{B}\big]\otimes\ket{0}_{\text{rest-of-chain}}.
\label{entang2}
\end{eqnarray}
  
\begin{figure}[ht!]
\centering
\resizebox{0.7\textwidth}{!}{
  \includegraphics{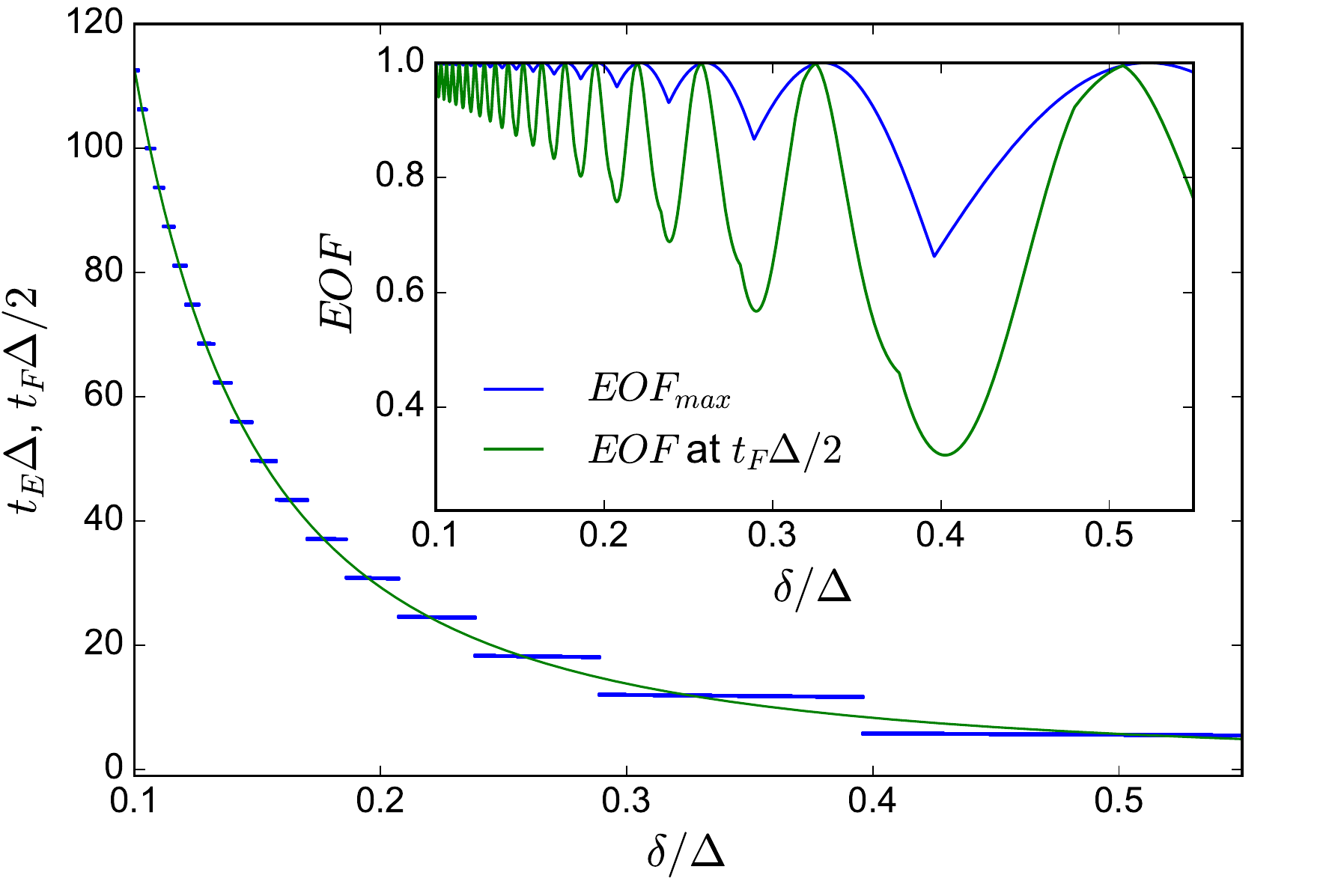}
}
\caption{Variation of $t_E$ with coupling ratio (blue segments) for the two-excitation protocol with overlaid analytical behaviour of $t_F/2$ (green line). Inset: maximum EOF at $t_E$ (blue line) and EOF at the analytical $t_F/2$ (green line).}
\label{maxEoFtwoex}       
\end{figure}

We now consider the analysis of the two-excitations protocol over a range of coupling ratios. Fig.\ \ref{maxEoFtwoex} shows the results for the same range of coupling ratios as that considered for the single-excitation protocol.
Considering the behaviour of $t_E$ first, we observe very similar behavior for both protocols, but we note that the intervals of almost constant $t_E$ fall at different ratios for the single- and two-excitation protocols. The intervals cover different ranges of ratios in both cases, and hence no protocol is faster overall. If a particular ratio is required, it would be necessary to compare the time-dependencies of both protocols as well as the corresponding $EOF_{max}$. 

The variation of $EOF_{max}$ with coupling ratio for the two-excitations case is shown in the inset of Fig.\ \ref{maxEoFtwoex}. We observe a similar oscillatory pattern to that of the single-excitation case, but now the maximum of $EOF_{max}$ within each arch does not decrease with the coupling ratio. As a result, it is possible to obtain EOF values extremely close to unity even for high ratios: a ratio as high as \TERL\ yields an $EOF_{max}$ greater than 0.999 and in just 5.57 time units. 

\begin{figure}[ht!]
\centering
\resizebox{0.7\textwidth}{!}{
  \includegraphics{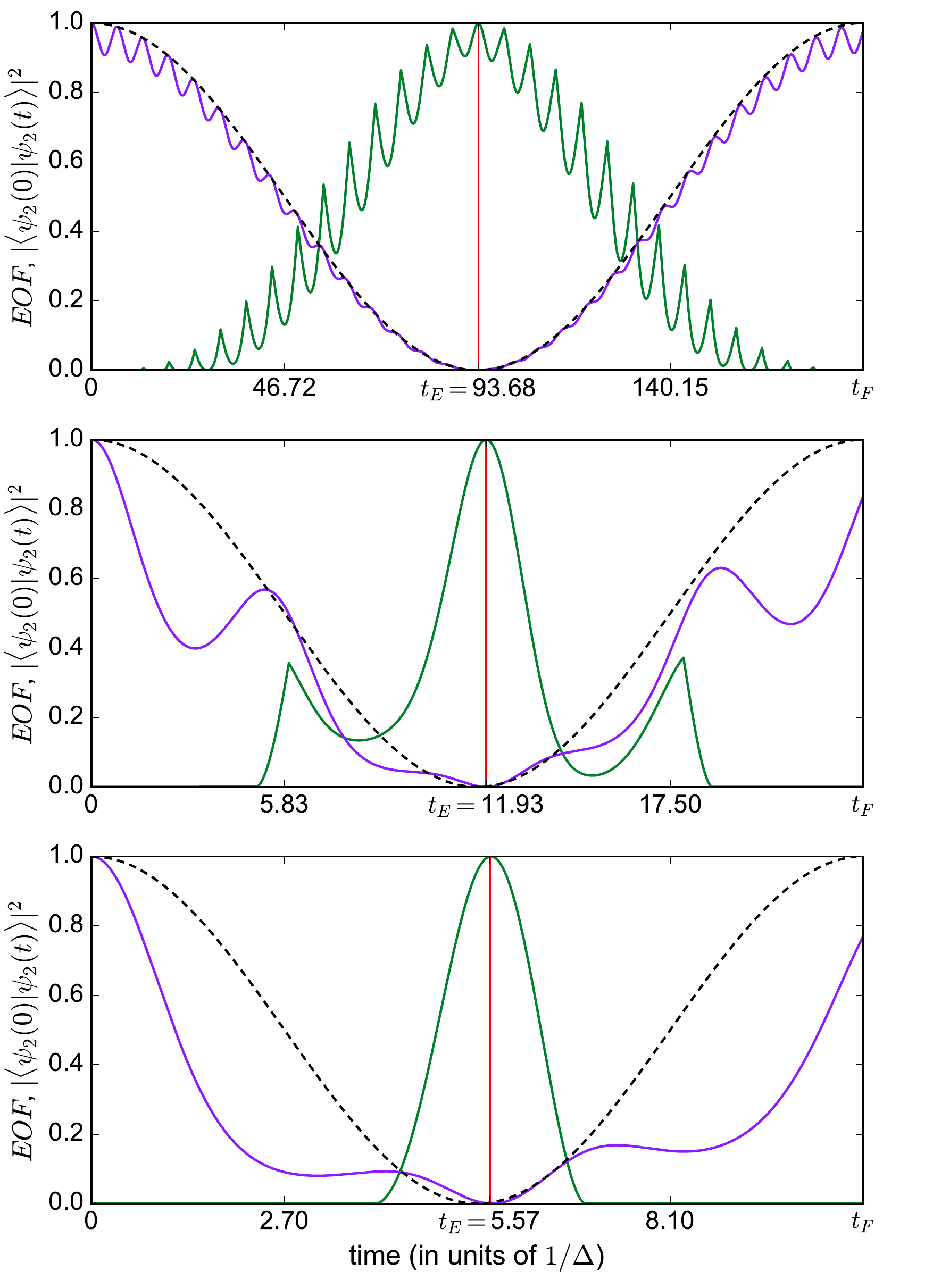}
}
\caption{Fidelity of the initial state (purple) and EOF (green) for the two-excitation protocol with ratios \TERF \ (top), \TERS \ (middle), and \TERL \ (bottom). The vertical red line denotes the time at which the maximum value of the EOF is obtained. The black dashed profile is the analytically obtained fidelity for each system.}
\label{panTwoEx}       
\end{figure}

As with the single-excitation case, we now consider in more detail the dynamics of the chain at three specific coupling ratios corresponding to maxima of $EOF_{max}$. We select coupling ratios of \TERF , \TERS , and \TERL, as shown in Fig.\ \ref{panTwoEx}. For each ratio, the value of $EOF_{max}$ is greater than 0.999. Similarly to the single-excitation protocol, we observe the maximum of EOF close $t_F/2$, where the fidelity reaches a value close to zero. However, for the two-excitations protocol, there is a larger difference between $t_F/2$ and $t_E$. Similarly we observe in the inset of Fig.\ \ref{maxEoFtwoex} a small shift between the local maxima of $EOF_{max}$ and EOF at $t_F/2$. Once more there is the  presence of the secondary oscillation in both the EOF and fidelity curves (Fig.\ \ref{panTwoEx}). However, the secondary oscillations of both fidelity and EOF present, in general, greater amplitudes, and the EOF curve presents cusps, a qualitatively different behaviour from the single excitation protocol. We attribute this to a more complex states interplay in the wave function which is more poorly approximated by the trimer toy model. For example, as the coupling ratio increases, the additional excitation in the chain center spreads over more central sites at $EOF_{max}$, so additional components should be added to Eq.\ (\ref{entang2}), which then takes the form
\begin{eqnarray}
\ket{\psi_2(t_E)}=\frac{1}{\sqrt{2}}\big(\ket{10}_\text{AC} + \ket{01}_\text{AC}\big)\otimes \big[\alpha\ket{1}_\text{B}\otimes\ket{0}_{\text{rest-of-chain}}+\beta\ket{0}_\text{B}\otimes\ket{\chi}_{\text{rest-of-chain}}\big],
\label{entang3}
\end{eqnarray}
a feature which cannot be reproduced by the trimer model.

\begin{figure*}[ht!]
\centering
\resizebox{1.0\textwidth}{!}{
  \includegraphics{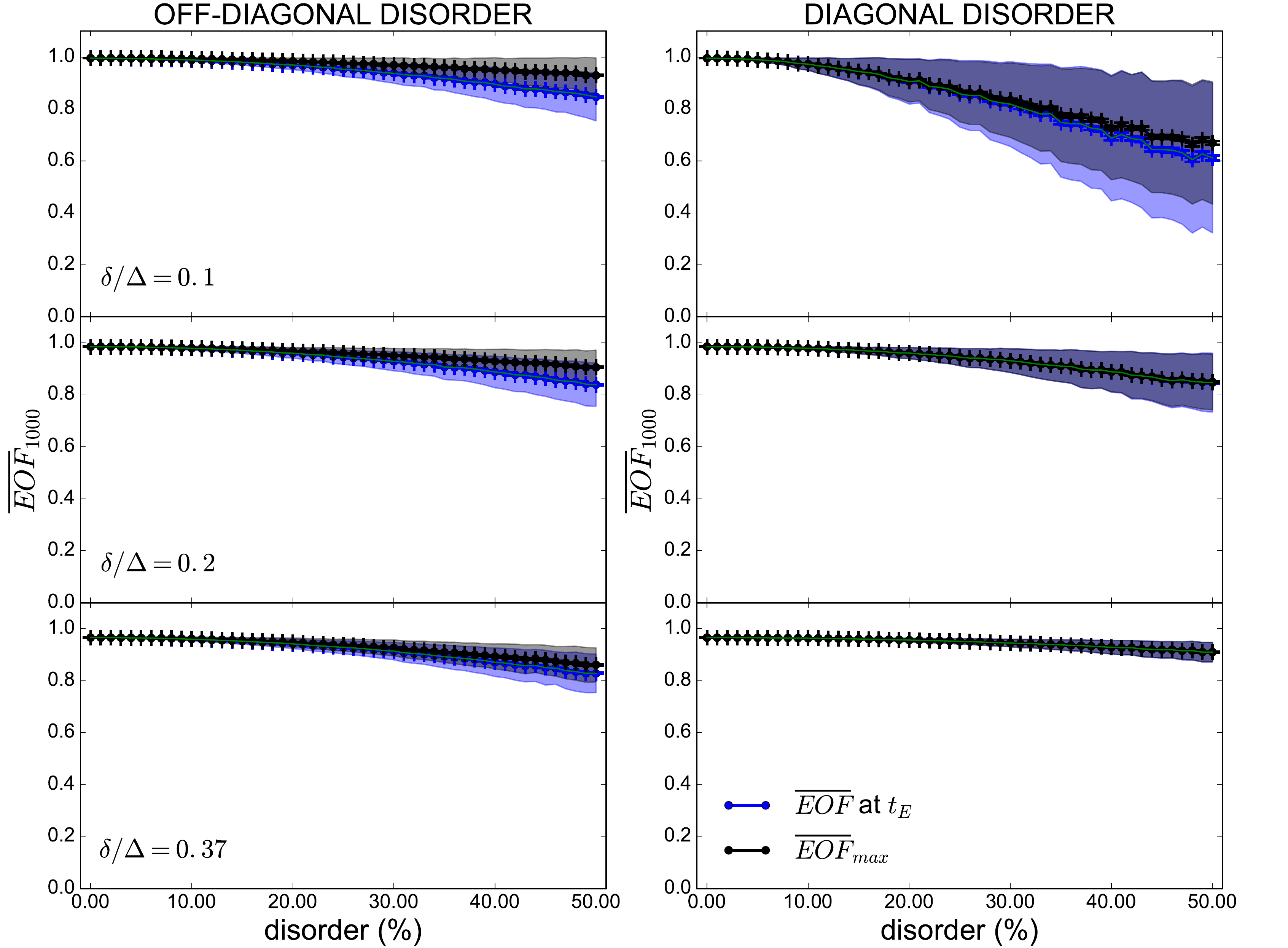}
}
\caption{Single-excitation protocol with coupling ratios of \SERF \ (top), \SERS \ (middle), and \SERL \ (bottom): averaged EOF at $t_E$ (blue dots) and averaged maximum EOF over windows of $t_F$ time units (black dots) for different levels of off-diagonal (left) and diagonal (right) disorders weighted against the weak coupling $\delta$. Grey and blue shadows represent the standard deviation, black and blue bars represent the standard error of the mean.}
\label{robGroup1}       
\end{figure*}

\begin{figure*}[ht!]
\centering
\resizebox{1.0\textwidth}{!}{
  \includegraphics{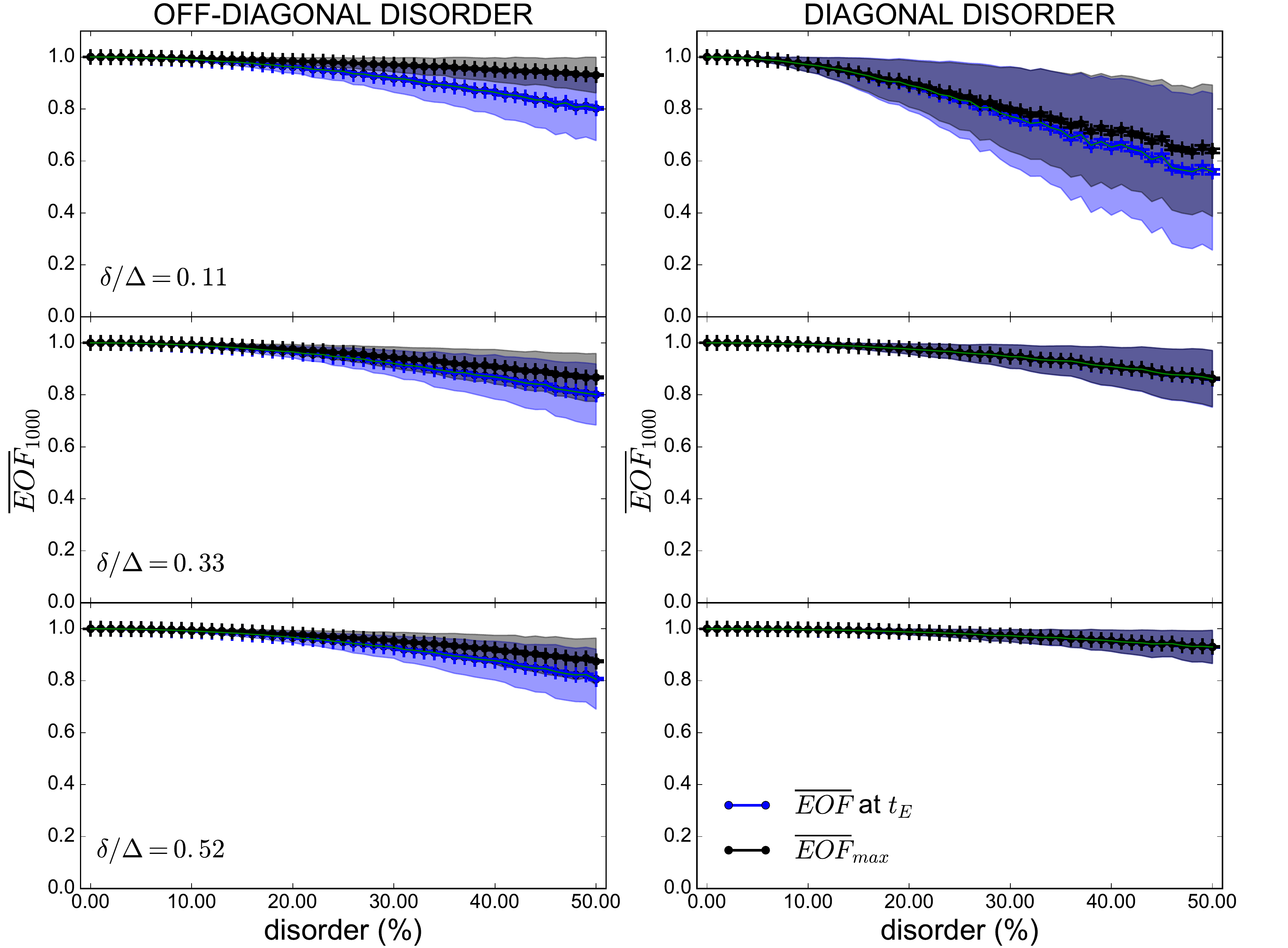}
}
\caption{Two-excitation protocol with coupling ratios of \TERF \ (top), \TERS \ (middle), and \TERL \ (bottom): averaged EOF at $t_E$ (blue dots) and averaged maximum EOF over windows of $t_F$ time units (black dots) for different levels of off-diagonal (left) and diagonal (right) disorders weighted against the weak coupling $\delta$. Grey and blue shadows represent the standard deviation, black and blue bars represent the standard error of the mean.}
\label{robGroup2}       
\end{figure*}

\subsection{Robustness to disorder}

The previous analysis has considered only ideal spin chains, however imperfections are likely to affect real physical devices. We will therefore consider the effect of two types of disorder on the model system, which can account for local, static perturbations and fabrication imperfections. These two types of disorder are modeled by introducing random terms on the off-diagonal (emulating imperfections of the couplings) and on the diagonal  (corresponding to fluctuations of the site energies) terms of the Hamiltonian.

We first consider the presence of off-diagonal disorder through the modification of our system couplings, $J_{i, i+1}$, such that $J_{i,i+1}^{\text{eff}}=J_{i,i+1}+E_Jd_{i, i+1}\delta$ with $J_{i,i+1}\in (\Delta, \delta)$. $E_J$ is a dimensionless scaling factor which sets the strength of the disorder and we weight the disorder with the weak coupling strength ($\delta$). To simulate the random nature of the disorder we introduce the pseudo-random number $d_{i, i+1}$, where $-0.5\leq d_{i, i+1}\leq 0.5$. 

Secondly, we consider the effect of diagonal disorder through the addition of a term  of the form $\epsilon_i = E_\epsilon d_i\delta$ to each diagonal element of the system Hamiltonian. Here $d_i$ is another pseudo-random number, with the same properties as the first. Again we use a dimensionless scaling factor $E_\epsilon$ and weight the disorder with the weak coupling strength. 

In analyzing the robustness of our protocols two different scenarios have been considered. In the first we calculate the average EOF over many disorder realisations, $\overline{EOF}$, at the time $t_E$ of maximum entanglement, of the unperturbed system. However, we note that due to the possibility of the perturbations shifting the time at which the EOF peak appears, we are no longer guaranteed to observe the maximum EOF at this time. It is therefore useful to also consider the average of the maximum EOF within the full time window in the presence of perturbations, $\overline{EOF}_{max}$, which is a measure of considerable practicality when calibration of the device is feasible. For both cases we present an (ensemble) average over 1000 realisations and consider a range of disorder weighted up to 50\% of the weak coupling strength.

We focus first on the single-excitation protocol, for each of the coupling ratios of Fig.\ \ref{panOneEx}, \SERF, \SERS, and \SERL. The results for off-diagonal and diagonal disorder are shown in the l.h.s and r.h.s panels of Fig.\ \ref{robGroup1}, respectively. It can be seen that, for each of the coupling ratios, the robustness against off-diagonal disorder is extremely high, with both $\overline{EOF}$ and $\overline{EOF}_{max}$ remaining always above 0.8. The $\overline{EOF}_{max}$ is almost independent of the coupling ratio, whereas the robustness of $\overline{EOF}$ is slightly diminished as the coupling ratio increases. Looking now at the blue and grey shadows, which represent the standard deviation of each ensemble, we observe that, for each ratio, the standard deviation of the single realisation is very small. For diagonal disorder, we observe a more pronounced dependence of the robustness of the protocol on the coupling ratio. For the smallest ratio of \SERF, both $\overline{EOF}$ and $\overline{EOF}_{max}$ are close to 0.6 at the maximum disorder strength. This is in stark contrast to the case for a coupling ratio of \SERL, in which both average EOF are close to 0.9 at the same relative disorder strength. Similarly, the standard deviation of the single realisation significantly decreases with increasing coupling ratio. In this case the system is more robust to diagonal disorder than off-diagonal disorder, but neither cause a reduction greater than 20\% of the maximum EOF value attainable with the unperturbed system.

Moving now to the two-excitation case, our results show that the trends in the results are similar, as shown in Fig.\ \ref{robGroup2}. We again observe a very high robustness against off-diagonal disorder with average EOF values never below 0.8, and $\overline{EOF}_{max}$ almost independent of the coupling ratio. As the coupling ratio increases, more robust becomes the system against diagonal disorder. In particular with coupling ratios of \TERS\ and \TERL\, we see that the $\overline{EOF}_{max}$ and $\overline{EOF}$ are basically indistinguishable and both remain above 0.85 for the former and 0.93 for the latter, even at the highest levels of disorder. We note that the standard deviation is slightly larger in general for the two-excitation protocol when compared to the single-excitation counterpart remaining very small at relatievly large coupling ratios. However, for the largest ratio the maximum entanglement is gained almost twice as fast as with the highest ratio considered with the single-excitation protocol. 

\subsection{Comparison of the entangling time scale and decoherence times for potential physical implementations}

We wish to qualitatively assess the possibility of performing our protocol on few experimental platforms. In order to do so, we consider some of the typical characteristic coupling energies for electron qubits in GaAs/AlGaAs quantum dots \cite{niko2004}, exciton qubits in self-assembled quantum dots \cite{damico2006,damico2002}, trapped ions \cite{muller2008} and superconducting qubits \cite{Houck2012,Gambetta2017}. We use this energies to estimate $\Delta$ and then $t_E$ for our two excitation protocol. Table \ref{times} shows results for the mid coupling ratio of 0.33 which though providing excellent fidelity and robustness, does no correspond to the best case scenario. From our numerics for the two excitation protocol with a ratio of 0.33 we have $t_E=\text{time}=11.93/\Delta$. We compare our results for the $t_E$ with typical decoherence times for the various hardwares. Table \ref{times} shows that for most platforms $t_E$ is much smaller than characteristic decoherence time, giving clear optimism on the possibility of using our protocol in experiments.

\begin{table}[h]
\centering
\caption{Approximated entangling times of our protocol for different experimental realisations and their decoherence times.}
\label{times}
\scalebox{0.85}{
\begin{tabular}{llll}
\textbf{Platform}       & \textbf{Characteristic energy, $\Delta$} & \textbf{Entangling time, $t_E$} & \textbf{Decoherence time} \\
QDs (electrons)       & \SI{0.05}{\milli\electronvolt}                 & \SI{20}{\nano\s}                         & \SI{1}{\micro\s}  \\
QDs (excitons)       & \SI{1}{\milli\electronvolt}                 & \SI{8}{\pico\s}                         & \SI{1}{\nano\s}  \\
trapped Rydberg Ca$^+$ ions & \SI{500}{\planckbar\mega\hertz}       & \SI{0.2}{\nano\s} &  \SI{10}{\micro\s} \\
Superconducting qubits  & \SI{1}{\giga\hertz}                   & \SI{0.1}{\nano\s}                     &  \SI{100}{\micro\s} 
%
%
\end{tabular}}
\end{table}

\section{\label{sec:concl}Conclusions}

We have demonstrated the potential for ABC spin chains to generate EPR states through two different entangling protocols. Importantly, our results prove the possibility of using a large range of characteristic coupling ratios for rapid generation of entangled states to a high fidelity. By increasing the value of $\delta$, the entangled state is generated faster and with only a very small-to-insignificant fidelity loss. For the single-excitation protocol and the highest ratio considered ($\delta/\Delta=\SERL$), we incur a very small reduction in EOF of approximately 0.05; on the other hand, for the two-excitation protocol, we can retain an EOF of almost unity for all characteristic coupling ratios we considered. This result is particularly encouraging due to the exponential speedup of both entanglement protocols that we observe as the coupling ratio increases. Furthermore, we have demonstrated excellent robustness to both diagonal and off-diagonal disorder for  these characteristic coupling ratios: not only is the robustness to off-diagonal disorder almost independent of coupling ratio, but that the robustness to diagonal disorder increases significantly as the coupling ratio is increased. This result exposes an interdependent relationship between the speed and robustness to disorder of both protocols. The ability to maximize both of these factors at little to no reduction in the maximum EOF allows us to propose ABC chains as rapid and reliable entanglement generation devices that present several advantages in front of previous implementations.
\par
First, our proposed device has the advantage of using static qubits (or `stationary qubits') without the need of moving entities. This gives ABC-type chains the potential to fulfill the roll of gates between shortly distanced quantum registers. Second, this protocol is driven by the natural dynamics of the chain limiting the need of user interaction with the system. This interaction however will not be null as initialization and extraction of the desired input/output states will be needed. However, this leads to an additional advantage related to the flexibility of our protocol: we can generate entangled states from different initial state injections, each of which might be convenient in different hardware implementations. Additionally, as shown in previous work \cite{estarellas2017robust}, this protocol also offers the possibility of localizing and therefore storing the entangled state in a topologically protected eigenstate. Robustness is another plus; as the resilience of the protocol against fabrication defects and time delays is highly favorable for real applications. The structure of the chain also allows for modification and even optimization as we have shown by trying to find a compromise between the fidelity of the desired entangled state, the overall robustness of the protocol and, rather importantly, the time needed for the entanglement operation to settle. These times are very relevant features of such devices due to the necessity to have quickly operating gates able to perform as many operations as possible before the coherence of the qubits in the implementation falls off. In that line, and after performing such optimization, our rescaled operation times offer feasible results under real experimental implementations, as illustrated in Table \ref{times}. All these points offer clear optimism in the prospective use of such devices in  real applications. 

\nonumsection{References}
\bibliography{papers_bib}
\bibliographystyle{unsrt}

\appendix{ - Derivation of $t_F$}
\label{appa}
The time needed for the trimer to retrieve its initial state ($t_F$), or fidelity period, can be obtained from the wavefunction, $|\Psi(t)\rangle$, of the system. We can write the initial state in terms of the eigenstates (Eq.\ (\ref{eigenstates})) and each of them can be time evolved through its propagator ($e^{-iEt}$):

\begin{equation}
|\Psi(t)\rangle=\frac{1}{\sqrt{2}}(a_{+}e^{-iE_{+}t}|\phi_{+}\rangle+a_{-}e^{-iE_{-}t}|\phi_{-}\rangle),
\end{equation}

where $a_{-}$ and $a_{+}$ are the coefficients resulting of mapping the initial state from the site basis (Eq.\ (\ref{is1}) or Eq.\ (\ref{is2}), as the two initial states are equivalent due to symmetry of the trimer model) into the eigenstate basis (Eq.\ (\ref{eigenstates})), 

\begin{equation}
	\begin{matrix}
a_{-}=\langle\Psi(0)|\phi_{-}\rangle \\
a_{+}=\langle\Psi(0)|\phi_{+}\rangle. 
	\end{matrix}
\label{coefficients}
\end{equation}

The fidelity time for will be equal to the time needed for all the propagators of the eigenstates to be equal to 1. Such time is found solving the following system of equations,

\begin{equation}
\begin{cases} 
e^{-i\sqrt{2}\eta t_F}=1\\ 
e^{i\sqrt{2}\eta t_F}=1, 
\end{cases}
\end{equation}

being the eigenvalues $E_{+}=\sqrt{2}\eta$ and $E_{-}=-\sqrt{2}\eta$. Thee fidelity time is therefore resolved to be $t_F=\pi/\sqrt{2}\eta$.

\appendix{ - Derivation of $\eta$}
\label{appb}
When calculating the single excitation spectrum of the 7 sites ABC chain seven different energy states are obtained. Two on an upper band $E^{+2}$, $E^{+3}$, two on a lower band $E^{-2}$, $E^{-3}$ and three sitting in between (energy gap), $E^{+1}$, $E^0$ and $E^{-1}$. From diagonalizing the full hamiltonian in terms of $\Delta$ and $\delta$ we obtain the following analytical forms of such eigenvalues:

\begin{equation*}
E^{\pm3}=\pm\frac{\sqrt{\Delta^2+3\delta^2+\sqrt{\Delta^4+6\Delta^2\delta^2+\delta^4}}}{\sqrt{2}}
\end{equation*}
\begin{equation*}
E^{\pm2}=\pm\sqrt{\Delta^2+\delta^2}
\end{equation*}
\begin{equation*}
E^{\pm1}=\pm\frac{\sqrt{\Delta^2+3\delta^2-\sqrt{\Delta^4+6\Delta^2\delta^2+\delta^4}}}{\sqrt{2}}
\end{equation*}
\begin{equation*}
E^0=0
\end{equation*}

We know we can relate the three states sitting in the energy band ($E^{\pm1},E^0$) to the trimer model. First we show that the  energies obtained from diagonalization of Eq.\ (\ref{hamiT}) are equivalent to the energies of an excitation sitting at sites A, B and C. We can rewrite our one excitation basis vectors in terms of the eigenvectors of the trimer (Eq.\ (\ref{eigenstates})) yielding to the following energy values,

	\begin{eqnarray}
    \left(
    \begin{matrix}
	1\\0\\0
	\end{matrix}
	\right)&=&\frac{1}{2}(\sqrt{2}|\phi_{0}\rangle+|\phi_{+}\rangle-|\phi_{-}\rangle)\rightarrow 
    \nonumber\\
& & E_A=\frac{1}{2}(\sqrt{2}E_{0}+E_{+}-E_{-})=\frac{1}{2}(\sqrt{2}·0+\sqrt{2}\eta+\sqrt{2}\eta)=\sqrt{2}\eta
	\label{eigenstates1}
\\
    \left(
    \begin{matrix}
	0\\1\\0
	\end{matrix}
	\right)&=&\frac{1}{\sqrt{2}}(|\phi_{+}\rangle+|\phi_{-}\rangle)\rightarrow \nonumber\\
& & E_B=\frac{1}{\sqrt{2}}(E_{+}+E_{-})=\frac{1}{\sqrt{2}}(\sqrt{2}\eta-\sqrt{2}\eta)=0
	\label{eigenstates2}
\\
	\left(
    \begin{matrix}
	0\\0\\1
	\end{matrix}
	\right)&=&\frac{-1}{2}(\sqrt{2}|\phi_{0}\rangle-|\phi_{+}\rangle+|\phi_{-}\rangle)\rightarrow \nonumber\\
& & E_C=\frac{-1}{2}(\sqrt{2}E_{0}-E_{+}+E_{-})=\frac{-1}{2}(-\sqrt{2}\eta-\sqrt{2}\eta)=-\sqrt{2}\eta.
	\label{eigenstates3}
	\end{eqnarray}

From this we can assume that the presence of the dimers in between are the responsible for the upper/lower bands. Therefore, the `effective' coupling between A-B and B-C, $\eta$, will be related to the energy difference between these sites. Such difference is $\Delta E=E_A-E_B=\sqrt{2}\eta$ . If we retrieve the energy obtained from the diagonalization of the full Hamiltonian $E^{\pm 1}$ we can finally obtain $\eta$,

\begin{equation}
\Delta E=|E^{+1}-E^{0}|=|E^{0}-E^{-1}|=\sqrt{2}\eta 
\end{equation}

\begin{eqnarray}
E^{+1}&=&\sqrt{2}\eta \rightarrow \frac{\sqrt{\Delta^2+3\delta^2-\sqrt{\Delta^4+6\Delta^2\delta^2+\delta^4}}}{\sqrt{2}}=\sqrt{2}\eta
\nonumber \\
& & \eta=\frac{\sqrt{\Delta^2+3\delta^2-\sqrt{\Delta^4+6\Delta^2\delta^2+\delta^4}}}{2}.
\end{eqnarray}

After rearranging in terms of $\delta/\Delta$ we obtain the expression presented in Eq.\ (\ref{eta}).

\end{document}